\documentclass[iop]{emulateapj}

\usepackage{epsfig, amsmath}

\def\MSUN   {\,$\hbox{M}_\odot$}

\def\NTHP   {N$_2$H$^+$}
\def\NTDP   {N$_2$D$^+$}
\def\HCOP   {HCO$^+$}
\def\HTCOP  {H$^{13}$CO$^+$}
\def\kms    {km\,s$^{-1}$}

\def\VINOUT {V$_{\rm in/out}$}
\def\JEANS  {$M_{\rm J}$}

\shorttitle{Infall in Starless Cores}
\shortauthors{Schnee et al.}

\begin{document}

\title{Correlating Infall with Deuterium Fractionation in Dense Cores}

\author{Scott Schnee\altaffilmark{1}, Nathan
  Brunetti\altaffilmark{1}, James Di Francesco\altaffilmark{2,3},
  Paola Caselli\altaffilmark{4}, Rachel Friesen\altaffilmark{1,5},
  Doug Johnstone\altaffilmark{2,3,6}, Andy Pon\altaffilmark{2,3,4}}

\email{sschnee@nrao.edu}

\altaffiltext{1}{National Radio Astronomy Observatory, 520 Edgemont
  Road, Charlottesville, VA 22903, USA}
\altaffiltext{2}{National Research Council Canada, Herzberg Institute
  of Astrophysics, 5071 West Saanich Road Victoria, BC V9E 2E7,
  Canada}
\altaffiltext{3}{Department of Physics \& Astronomy, University of
  Victoria, PO Box 3055 STN CSC, Victoria, BC, V8W 3P6, Canada}
\altaffiltext{4}{School of Physics and Astronomy, University of Leeds,
  Leeds LS2 9JT, UK}
\altaffiltext{5}{Dunlap Institute for Astronomy and Astrophysics,
  University of Toronto, 50 St. George Street, Toronto M5S 3H4,
  Ontario, Canada}
\altaffiltext{6}{Joint Astronomy Centre, 660 North A'ohoku Place,
  University Park, Hilo, HI 96720, USA}

\begin{abstract}

We present a survey of \HCOP\ (3-2) observations pointed towards dense
cores with previous measurements of $N$(\NTDP)/$N$(\NTHP).  Of the 26
cores in this survey, five show the spectroscopic signature of
outward motion, nine exhibit neither inward nor outward motion,
eleven appear to be infalling, and one is not detected.  We compare
the degree of deuterium fractionation with infall velocities
calculated from the \HCOP\ spectra and find that those cores with
[D]/[H] $>$ 0.1 are more likely to have the signature of inward
motions than cores with smaller [D]/[H] ratios.  Infall motions are
also much more common in cores with masses exceeding their thermal
Jeans masses.  The fastest infall velocity measured belongs to one of
the two protostellar cores in our survey, L1521F, and the observed
motions are typically on the order of the sound speed.

\end{abstract}

\keywords{stars: formation,  Stars: protostars}

\section{Introduction}

Starless cores are dense (n $\sim$ 10$^5$\,cm$^{-3}$), cold (T
$\sim$10\,K), and compact (r $\sim$ 0.05\,pc) objects embedded within
molecular clouds \citep{DiFrancesco07}.  Starless cores are the
transition phase between the comparatively diffuse molecular cloud
material and a new generation of protostars. The cores are often
identified from their (sub)millimeter dust continuum emission and
classified as starless from the lack of infrared emission indicative
of an embedded protostar \citep[e.g.,][]{Nutter07, Hatchell07,
  Enoch08}.  Recent (sub)millimeter interferometric observations have
shown that low-luminosity protostars have been overlooked by {\it
  Spitzer} in roughly 10\% of supposedly starless cores
\citep{Schnee12}.

In the cold and dense interiors of starless cores, chemical reactions
lead to deuterium enrichment of some molecules that remain in the gas
phase, such as \NTHP \citep[see, e.g.,][]{Caselli08}.  In a survey of
low-mass cores in nearby molecular clouds, \citet{Crapsi05} measured
$N$(\NTDP)/$N$(\NTHP) values ranging from 0.01 to 0.44, much larger
than the [D]/[H] elemental abundance of $\simeq 1.5 \times 10^{-5}$
\citep{Oliveira03} found in the ISM.  \citet{Crapsi05} concluded that
the $N$(\NTDP)/$N$(\NTHP) ratio is an indicator of core evolution such
that a high ratio implies that a core is close to transitioning from
being starless to forming a protostar.  \citet{Friesen13} showed that
the deuterium fraction in cores in the Perseus molecular cloud tends
to increase with H$_2$ column density and with increasing core
concentration, in agreement with the Crapsi results.
\citet{Emprechtinger09} found that the \NTDP/\NTHP\ ratio is also an
indicator of evolutionary state for Class 0 protostars, with the
youngest protostars having the highest deuterium fractionation and
clearest signs of ongoing collapse.  In the Perseus sample, the most
highly deuterated protostellar cores are still associated with
significant envelope mass (Friesen et al.~2013).  Therefore, in the
evolution of a collapsing core, the $N$(\NTDP)/$N$(\NTHP) ratio is
expected to peak just as the protostar is formed.  In a sample of
starless cores, one might expect the \NTDP/\NTHP\ ratio to be roughly
indicative of the likelihood of collapse, with scatter due to
differences in core properties such as mass, temperature, non-thermal
motions, and magnetic field strength.

Infall and outflow motions in dense cores are revealed by asymmetries
in their spectral line profiles.  In a simple model of collapse,
density, optical depth, and excitation temperature increase towards
the center of the core.  Inward motions of the side of the core nearer
to the observer induce a redshift relative to the average core
velocity, and, as a result, one expects to see red-shifted
self-absorption of optically thick lines.  \citet{Myers96} and
\citet{DeVries05} present sets of equations that can be used to derive
infall velocity in idealized cases of core velocity and density
profiles, and these models match observations with reasonable
accuracy.  In a real core, the inferred infall velocity depends on the
point at which the observed line becomes optically thick and the
variation of the infall velocity with radius, complicating the
interpretation of the observed spectra.  Nevertheless, surveys of
infall motions in dense cores have been succesfully carried out using
various molecular tracers, such as \HCOP\ \citep{Gregersen00}, HCN
\citep{Sohn07}, CS \citep{Williams99, Lee99, Lee01}, and
\NTHP\ \citep{Williams99, Lee99, Lee01, Crapsi05}.

Infall in a starless core will result in a localized density increase
due to mass accumulation and a decrease in the central core
temperature due to the additional shielding from the interstellar
radiation field.  These conditions promote increased deuteration, so
one might expect to see large infall velocities in those cores with
the highest degree of deuteration.  To test the hypothesis of a
correlation between chemical and dynamical evolution, we observed the
\HCOP\ (3-2) and \HTCOP\ (3-2) lines toward every core in the
\citet{Crapsi05} sample for which the $N$(\NTDP)/$N$(\NTHP) ratio was
measured.  If those cores with the highest $N$(\NTDP)/$N$(\NTHP) ratio
are really on the verge of forming a protostar, then we expect them to
also show infall signatures.  Those cores with smaller observed
deuterium enrichment would have slower, or non-existent, infall
motions.

Similarly, those cores more massive than their thermal Jeans mass
might also be expected to exhibit infall motions.  We also compare the
observed motions with the Jeans stability of each core to test the
hypothesis that those cores with $M$/\JEANS\ $>$ 1 will have inward
motions while cores with $M$/\JEANS\ $<$ 1 will appear to be static.
More details on this survey are reported below.

\section{Observations} \label{OBSERVATIONS}

The 26 dense cores in this survey are those for which the
$N$(\NTDP)/$N$(\NTHP) ratios were measured by \citet{Crapsi05}.  The
\citet{Crapsi05} sample is meant to be representative of evolved
(i.e.~dense) starless cores, with their high densities derived from
\NTHP\ (1-0) or 1.2\,mm dust continuum emission.  Although thought to
be starless by \citet{Crapsi05}, the cores L1521F \citep{Bourke06} and
L328 \citep{Lee09} were later found to be protostellar.  The
\citet{Crapsi05} cores are nearby, i.e., within 250\,pc, making them
comparatively easy to study.  We chose to observe \HCOP\ (3-2) to
trace inward and outward motions in these dense cores.  \HCOP\ (3-2)
was shown by \citet{Gregersen00} to be a good tracer of line
asymmetries in starless cores.  The effective critical density of
\HCOP\ (3-2), 6.3$\times$10$^4$\,cm$^{-3}$ \citep{Evans99}, is
well-matched to the densities of the cores in the \citet{Crapsi05}
sample, 8$\times$10$^4$\,cm$^{-3}$ - 1.4$\times$10$^6$\,cm$^{-3}$.
The observed spectra are likely to be affected by optical depth and
depletion such that \HCOP\ (3-2) may not be tracing exactly the same
portion of every core in our sample.  Although this source of
uncertainty is unavoidable in a survey such as ours, we outline a
possible path around this uncertainty in Section \ref{FUTURE}.

Our observations were centered on the position of the peak
$N$(\NTDP)/$N$(\NTHP) given in \citet{Crapsi05}.  The [\NTDP]/[\NTHP]
ratio is expected to peak in the region with the highest density and
coldest temperature.  High densities and low temperatures (and
therefore low thermal support) are conditions that are likely to lead
to gravitational collapse.  The coordinates of our pointed
observations and the degree of deuteration are given in Tables
\ref{OBSTABLE} and \ref{PROPTABLE}.  The noise values of the spectra
are given in Table \ref{OBSTABLE}.  The observed \HCOP\ (3-2) spectra
are shown in Figures \ref{SPECTRA1A} - \ref{SPECTRA1D}.  

The data were obtained at the James Clerk Maxwell
Telescope\footnote{The James Clerk Maxwell Telescope is operated by
  the Joint Astronomy Centre on behalf of the Science and Technology
  Facilities Council of the UK, the Netherlands Association for
  Scientific Research, and the National Research Council of Canada}
(JCMT) using the 1\,mm Receiver RxA3 front-end and the ACSIS backend
throughout semesters 10A and 10B.  Each core was observed in
position-switching mode with an off-position located (-300,-300)
arcseconds (in the J2000 coordinate frame) from the core position (see
Table 1).  ACSIS was configured to cover two 250\,MHz-wide windows,
one centered on \HCOP\ (3-2) at 267.558\,GHz and another centered on
\HTCOP\ (3-2) at 260.255\,GHz.  Each window was divided into 8192
channels for a resolution of $\sim$0.05\,\kms.  Each core was observed
for about 45 minutes of integration and the beam size of the
observations was $\sim$20\arcsec\ FWHM.  The JCMT beam size is fairly
well matched to the beam sizes of the \NTHP\ (1-0) and \NTDP\ (2-1)
observations presented in \citet{Crapsi05}, i.e., 26\arcsec\ and
16\arcsec\ FWHM, respectively.

After acquisition, the data were reduced using the Starlink package.
Each integration was first visually checked for baseline ripples and
extremely large spikes, and such data were removed from the ensemble,
if necessary.  Spectral baselines were subtracted, frequency axes
converted to velocities, and spectra edges trimmed using Starlink
scripts kindly provided by T. van Kempen.  The integrations were
visually inspected and written out as FITS format files using other
standard Starlink routines.  The spectra for each core were co-added
outside of Starlink, using the IDL programming language.  Spectra were
converted to main beam brightness temperatures using an assumed
efficiency of
0.75\footnote{http://www.jach.hawaii.edu/JCMT/spectral\_line/Standards/beameff.html}.
\HCOP (3-2) emission was detected towards each core except for
L1495A-S, but \HTCOP\ (3-2) emission was only detected towards a few
cores (L1521F, Oph D, L1689B, and L328; see Figures
\ref{SPECTRA1A}-\ref{SPECTRA1D}).  Hence, the \HTCOP\ data will not
be discussed further in this paper.

\section{Results} \label{RESULTS}

The \HCOP\ spectra were fit with a Gaussian model as well as two
simple infall/outflow models, the ``two-layer'' model from
\citet{Myers96} and the ``HILL5'' model from \citet{DeVries05}.  The
two-layer model describes the case of two regions with differing
excitation temperatures (lower for the front, higher for the rear) and
identical velocity dispersions and total optical depth.  The HILL5
model describes the case of a core whose excitation temperature peaks
at the center and falls off linearly with radius.  The HILL5 core also
has its front and rear halves moving towards or away from each other.
Both models reproduce spectra with one peak and a shoulder reasonably
well, though the HILL5 model is better than the two-layer model at
fitting spectra with two peaks \citep{DeVries05}. The five free parameters
in both infall/outflow models are inward/outward velocity, kinetic
temperature, peak optical depth, velocity dispersion, and velocity
with respect to the local standard of rest.  The dynamic models assume that
the two peaks in the prototypical infall/outflow profile arise from
self-absorption and not from having two unrelated sources with
slightly different velocities along the same line of sight.  This
assumption can be tested through observations of an optically thin
tracer that should peak at the same velocity as the self-absorption
dip in the optically thick spectrum if no radial motion is present.
The \HTCOP\ (3-2) observations described in Section \ref{OBSERVATIONS}
were intended to be this optically thin tracer, but detections were
rare.  The \NTHP\ and \NTDP\ observations presented in
\citet{Crapsi05} are sufficient to show that there are no confounding
superpositions in our sample.  In the relatively few cases where
\HTCOP\ (3-2) is detected, the line peaks at the same velocity as the
\NTHP\ and \NTDP\ spectra.  The Gaussian and dynamic models were fit
with the MPFIT suite of functions \citep{Markwardt09}.

For each core, the Gaussian and dynamic model fits were compared with
F-tests, taking into account the chi-squared values for the fits and
the degrees of freedom in the fits.  For those cores where the F-test
prefers a dynamic model and the inward/outward motions are at least
three times the uncertainty of the motion, we treat the core as having
inward or outward motions for the rest of this paper.  We describe as
`static core' those cores for which the F-test prefers the Gaussian
model or the uncertainty in the radial motions is greater than three
times the magnitude of the motion.  The signal-to-noise ratio of the
L1495A-S observations is too low to attempt a fit to either model.
The peak intensities, integrated intensities, and velocity dispersions
(standard deviations, or $\sigma$ for a Gaussian velocity
distribution) measured from the best fits to the \HCOP\ spectra are
given in Table \ref{OBSTABLE}, with the intensities given on T$_{\rm
  mb}$ scale.  Note that the cores L1521F and L328 are actually
protostellar, not starless, and L1521F has the largest infall velocity
in our survey.  The infall velocity of L1521F was measured to be
0.2-0.3\,\kms\ by \citet{Onishi99}.  The presence of an embedded
protostar likely reduces the accuracy of the HILL5 and two-layer
models (note the relatively poor fit to the blue peak of the
spectrum), so we trust the detailed model of \citet{Onishi99} to
provide a more accurate estimate of the infall velocity than we derive
here.  The kinematic state and preferred model of each core as
determined by the model fits are given in Table \ref{VELOCITYTABLE},
along with previous kinematic classifications from the literature (see
\S \ref{DISCUSSION}).  In Table \ref{VELOCITYTABLE}, we also give the
velocity with respect to the local standard of rest from the best-fit
model to our \HCOP\ spectra , along with the LSR velocity fit to the
\NTHP\ (1-0) spectra from \citet{Crapsi05}.  We find that the median
of the absolute value of the velocity difference between the LSR
velocity from \HCOP\ and \NTHP\ is 0.07\,\kms\ with a standard
deviation of 0.1\,\kms.  This velocity difference likely comes from
velocity gradients along the line of sight due to inward/outward
motions, since \HCOP\ and \NTHP\ will sometimes probe different layers
within a core.  The uncertainties given in Table \ref{VELOCITYTABLE}
are the 1\,$\sigma$ standard deviations.

Masses and radii for the cores in this survey were taken from the
literature and calculated from (sub)millimeter continuum surveys.
Most core masses are derived from the observed flux at
850\,\micron\ or 1.2\,mm using the formula
\begin{equation} \label{MASSEQ}
\begin{split}
M = 0.12 \left[ {\rm e}^{\left(\frac{1}{1.439} \frac{\lambda}{\rm mm}
    \frac{T}{10\,{\rm K}} \right)^{-1}}-1 \right]
\left(\frac{\kappa_\nu}{0.01\,{\rm cm}^2\,{\rm g}^{-1}}\right)^{-1} \\
 \left(\frac{S_\nu}{{\rm Jy}}\right) \left(\frac{D}{100\,{\rm
    pc}}\right)^2 \left(\frac{\lambda}{{\rm mm}}\right)^3 {\rm
  M}_\odot .
\end{split}
\end{equation}

We assume that all cores are isothermal at 10\,K, as assumed in the
core mass calculations of \citet{Crapsi05} and consistent with the
average temperatures derived from NH$_3$ observations of starless
cores \citep{Tafalla02, Schnee09}.  Real cores have temperature
gradients ranging from $\sim$13\,K at large radii to $\sim$6\,K at the
center \citep[e.g.,][]{Crapsi07, Pagani07, Schnee07b}, but these
variations in temperature do not affect the conclusions of this paper.
We assume a value of $\kappa_\nu$ at 850\,\micron\ of 0.01\,cm$^2$
g$^{-1}$ and $\kappa_\nu$ at 1.2\,mm of 0.005\,cm$^2$ g$^{-1}$, as
assumed in \citet{Sadavoy10} and \citet{Crapsi05}, respectively.
These values for $\kappa_\nu$ are consistent with each other for an
emissivity spectral index of 2, in agreement with observations of the
starless cores L1498 \citep{Shirley05} and TMC-1C
\citep{Schnee10}. Distances ($D$) to the cores and references used to
determine core masses are given in Table \ref{PROPTABLE}.  We were
unable to find previously published values for the masses and radii of
the cores L1495 and L1400A.

Following \citet{Sadavoy10}, we calculate the Jeans mass for the cores
in this survey using the formula
\begin{equation} \label{JEANSEQ}
M_J = 1.9 \left(\frac{T}{10\,{\rm K}}\right)
\left(\frac{R}{0.07\,{\rm pc}}\right) {\rm M}_\odot .
\end{equation}
As in the calculation of core masses, when calculating the Jeans mass
we assume that all cores are isothermal at $T = 10$\,K and use the
radii ($R$) given in Table \ref{PROPTABLE}.

\section{Discussion} \label{DISCUSSION}

A plot of infall velocity (see Table \ref{VELOCITYTABLE}) vs.~degree
of deuteration (see Table \ref{PROPTABLE}) is shown in Figure
\ref{INFALLFIG}.  Nine cores show no evidence for infall or outflow,
five cores have outward motions, and eleven have inward motions.
Below an $N$(\NTDP)/$N$(\NTHP) ratio of 0.1, there is no correlation
with infall velocity.  For the eight cores with $N$(\NTDP)/$N$(\NTHP)
$\ge$ 0.1, six cores have infall velocities and two cores appear to 
be static.  We therefore find a general trend, though not a
quantitative correlation, that inward motions are more likely to be
found at higher levels of deuteration whereas at lower levels of
deuteration, inward and outward motions are roughly equally likely to
be found.  The \HCOP\ spectra therefore support the idea that the most
chemically evolved starless cores are also dynamically evolved.

\citet{Simpson11} found that the Jeans stability of a pre-stellar core
provides a reasonable predictor of its dynamical state.  As shown in
Figure \ref{INFALLFIG} (see also Tables \ref{PROPTABLE} and
\ref{VELOCITYTABLE}), we find that three Jeans-stable cores (those
with $M/M_{\rm Jeans} < 1$) show inward motions, two cores appear to
be static, and four show outflow motions.  Those cores with $M/M_{\rm
  Jeans} > 1$ are much more likely to have line asymmetries associated
with infall (eight cores) than outward motions (no cores), and six cores 
with $M/M_{\rm   Jeans} > 1$ are best fit by Gaussian profiles.  Our results 
are in agreement with those of \citet{Simpson11} in that the Jeans 
stability of dense cores is a useful indicator of likely infall 
candidates, and the identification of collapsing cores becomes more 
likely with increasing $M/M_{\rm Jeans}$.  As shown in Figure 
\ref{INFALLFIG}, those cores with $N$(\NTDP)/$N$(\NTHP) $>$ 0.1 also 
have $M/M_{\rm Jeans} > 1$.

Many cores with lower degrees of deuteration, however, also have high
$M/M_{\rm Jeans}$ (L1495A-S, TMC-1C, CB 23, L1517B, L1689B, L492, and L328) 
and these cores sometimes exhibit infall motions that are a consequence of
their Jeans instability.  There are no cores with $M/M_{\rm Jeans}$ $<$ 1 and
$N$(\NTDP)/$N$(\NTHP) $>$ 0.1.  Those cores with $M/M_{\rm Jeans} >
1$, but no indication of inward motions, may be supported by
mechanisms other than thermal pressure (e.g., turbulent or magnetic
pressure).  Alternatively, infall may indeed be on-going, but the
signature of this motion could be missing from the \HCOP\ (3-2)
spectra in some cores due to freeze-out \citep{Roberts10} or rotation
\citep{Redman04}.  We suggest that those cores with super-Jeans masses
but no indication for infall are a good sample for future
observations.

The origin of outward motions in five starless cores in our survey, as
indicated by negative infall velocities in Table \ref{VELOCITYTABLE}
and Figure \ref{INFALLFIG}, is not fully understood.  Such outward
motions have been seen in previous surveys \citep[e.g.,][]{Sohn07,
  Lee11} and may indicate oscillatory behavior \citep{Lada03,
  Broderick10}.  Indeed, oscillatory behavior may also be responsible 
for some of the inward motions observed in our \HCOP\ survey.  SPH 
simulations of nominally super-Jeans cores
modelled as Bonnor-Ebert spheres may oscillate rather than collapse if
the exterior temperatures of the cores are high enough
\citep{Anathpinkida13}. As the sometimes-different classifications of
cores as 'expanding', 'oscillating', or 'contracting' in Table
\ref{VELOCITYTABLE} make clear, observing cores at slightly different
positions with different molecular tracers can lead to very different
conclusions, an effect which surely adds considerable scatter to plots
like Figure \ref{INFALLFIG}.

Only three of the sixteen cores included in four surveys of inward and
outward motions \citep[][this work]{Crapsi05, Sohn07, Lee11} showed
the same behaviour in all cases (L1544, L694-2, and L1197, which all
show inward motions).  Infall in the cores L1544 and L694-2 has been
particularly well-studied, with evidence for extended infall at speeds
$\le$0.1\,\kms\ seen in \NTHP\ observations \citep{Tafalla98,
  Williams06, Keto10}. The lack of a 1:1 correlation between infall
velocity and high $N$(\NTDP)/$N$(\NTHP) or $M/M_{\rm Jeans}$ is likely
to be at least partly a result of the uncertainties in measuring the
dynamics of cores with a single molecular tracer at a single position.
In addition, because starless cores won't all reach exactly the same
level of $N$(\NTDP)/$N$(\NTHP) just before collapse, one would not
expect a perfect correlation with infall even in a comprehensive set
of observations.

It may seem surprising that some cores in this survey have significant
infall velocities without having similarly large $N$(\NTDP)/$N$(\NTHP)
ratios.  As mentioned above, in some cases the inward motions may be due 
to oscillations in a stable core rather than being caused by collapse, 
in which case the low $N$(\NTDP)/$N$(\NTHP) ratio is a true indicator 
of the stability of a core.  Alternatively, \HCOP\ may be preferentially 
tracing accretion motions (i.e., motions of material onto the core) 
instead of contraction or expansion motions of the core itself, as 
\HCOP\ is also very abundant at low densities (i.e., in the core envelope 
and the surrounding molecular cloud), unlike \NTHP.  For example, 
molecular abundances in the contracting core TMC-1C are affected by 
accretion from its environment, from which it is gaining material rich 
in \NTHP\ but likely not in \NTDP\ \citep{Schnee05, Schnee07a}.  It is 
also possible that cores with significant infall velocities but low levels 
of deuterium fractionation are too young to have significant deuterium
enrichment despite their high central densities.  For instance, the
cores L1521E \citep{Tafalla04} and L1689B \citep{Lee03} have been
identified as dense cores with little chemical evolution, likely due
to their relative youth ($\leq 1.5 \times 10^5$ yr).  In these
chemically young cores, where CO is not highly depleted, \HCOP\ may be
a better tracer of infall in the central high density region, compared
to chemically evolved cores.  Therefore, a mismatch between the
chemical and dynamical states of a starless core inferred by this
survey could be due to 1) a single \HCOP\ (3-2) spectrum being an
incomplete tracer of the dynamics of an oscillating dense core, 2) the
$N$(\NTDP)/$N$(\NTHP) ratio within a core being lowered by accretion
from the surrounding medium, or 3) the timescale of chemical evolution
being significantly longer than the dynamical timescale.

The magnitude of the inward and outward velocities are low, similar to
the sound speed ($\sim$0.2\,\kms\ for a 10\,K core). This result
suggests that the initial conditions from which the motions are
derived were near hydrostatic equilibrium.  That is, the infall
motions are not dominated by highly supersonic flows attributed to
cloud turbulence but more likely derive from gravitational collapse of
a core that has lost pressure support.  The outward motions observed
are consistent with either oscillations of a near gravitational
equilibrium object \citep{Broderick10, Anathpinkida13} or thermal
expansion of an over-pressured but gravitationally unbound core, as
would be expected for a core with $M/M_{\rm Jeans} \lesssim 1$.

\section{Future Work} \label{FUTURE}

As outlined in this paper, there is a need for a uniform
multidimensional study of the kinematics and chemistry of dense cores.
The work presented in this paper is limited by the single-pointing
observations.  Spectral line maps would be better able to determine
the location of peak infall or outflow, and to determine whether a
core shows infall or outflow spectral signatures consistently across
its extent.  Part of the difference between the infall surveys
presented in Table \ref{VELOCITYTABLE} can be attributed to the
different pointing centers of the observations.  To the extent that
the velocity field changes across a core, these positional differences
will confuse the conclusions of pointed surveys.  Furthermore, the
``center'' of a core is not always a well-defined location.  For
example the center of L183 is offset by $\sim$20\arcsec\ between the
studies of \citet{Crapsi05} and \citet{Pagani07}.  A second conclusion
that can be drawn from Table \ref{VELOCITYTABLE} is that there is no
single best molecular tracer of inward and outward motions in dense
cores.  Although \HCOP\ is relatively bright and shows infall motion
clearly in many cores, it may not trace the regions in starless cores
with the lowest temperatures, highest densities, and most depletion.
In this case, the infall velocity measured from \HCOP\ spectra is
likely to be an underestimate of the maximum infall speed.  To map
precisely the inward and/or outward motions in a core, one would want
to have several spectral maps with different molecules and rotational
transitions.  These observations could then be modeled with a
radiative transfer code to constrain self-consistently the density,
temperature, chemistry, and kinematic profiles of the cores.  Such a
survey would require a substantial amount of observing time.  We note
that the pointed observations of a single infall tracer in this paper
required over 30 hours of telescope time (including overheads) over
multiple semesters with the JCMT.  The combination of high sensitivity
with high spatial and spectral resolution provided by the Atacama
Large Millimeter/submillimeter Array (ALMA) may make it possible to
carry out a comprehensive survey of the connection between the
chemistry, kinematics, and role of environment in dense cores.

\section{Summary} \label{SUMMARY}

\citet{Crapsi05} predicted that dense cores with $N$(\NTDP)/$N$(\NTHP)
$>$ 0.1 will show evidence for infall.  To test this hypothesis, we
observed \HCOP\ (3-2) towards the \citet{Crapsi05} sample of 26 dense
cores (24 starless and 2 protostellar) and analyzed the spectra to
look for inward and outward motions.  We find that those cores with
the largest values of $N$(\NTDP)/$N$(\NTHP) and $M/M_{\rm Jeans}$ are
indeed more likely to show the signature of infall than the population
as a whole, though some cores that have lower levels of deuterium
fractionation and greater Jeans stability also show inward motions.
Since asymmetries from inward and outward motions vary by position
within cores \citep{Lee01,Lada03} and vary by tracer observed
\citep[e.g.,][]{Lee11}, it is not surprising that we do not see very
tight correlations between the degree of deuterium fractionation or
Jeans stability with infall velocity.  The presence of outward motions, 
and perhaps some of the inward motions, in these presumably long-lived 
cores is possibly a result of oscillations \citep{Lada03, Broderick10, 
Anathpinkida13} and such outward motions have been seen before in 
previous surveys of dense cores \citep[e.g.,][]{Sohn07,Lee11}.  The 
magnitudes of the inward and outward motions are on the order of the 
sound speed, suggesting that the cores formed near hydrostatic 
equilibrium rather than out of supersonic flows.

\acknowledgments

We thank Malcolm Currie for his help reducing JCMT spectra. We
thank our anonymous referee for suggestions that have significantly
improved the clarity and content of this paper.  The
National Radio Astronomy Observatory is a facility of the National
Science Foundation operated under cooperative agreement by Associated
Universities, Inc.  JDF and DJ acknowledge support by the National
Research Council of Canada and the Natural Sciences and Engineering
Council of Canada (NSERC) via a Discovery Grant.  AP is partially
supported by the NSERC graduate scholarship program.  RF is a Dunlap
Fellow at the Dunlap Institute for Astronomy and Astrophysics,
University of Toronto.  The Dunlap Institute is funded through an
endowment established by the David Dunlap family and the University of
Toronto.  The James Clerk Maxwell Telescope is operated by the Joint
Astronomy Centre on behalf of the Science and Technology Facilities
Council of the United Kingdom, the Netherlands Organisation for
Scientific Research, and the National Research Council of Canada.  The
JCMT program ID numbers associated with this project are M10AC002,
M10BC001, and M11BC003.  This research has made use of the SIMBAD
database, operated at CDS, Strasbourg, France.

{\it Facilities}: JCMT

{}

\renewcommand{\thefigure}{1\alph{figure}}
\begin{figure}
\epsscale{1.0} 
\plotone{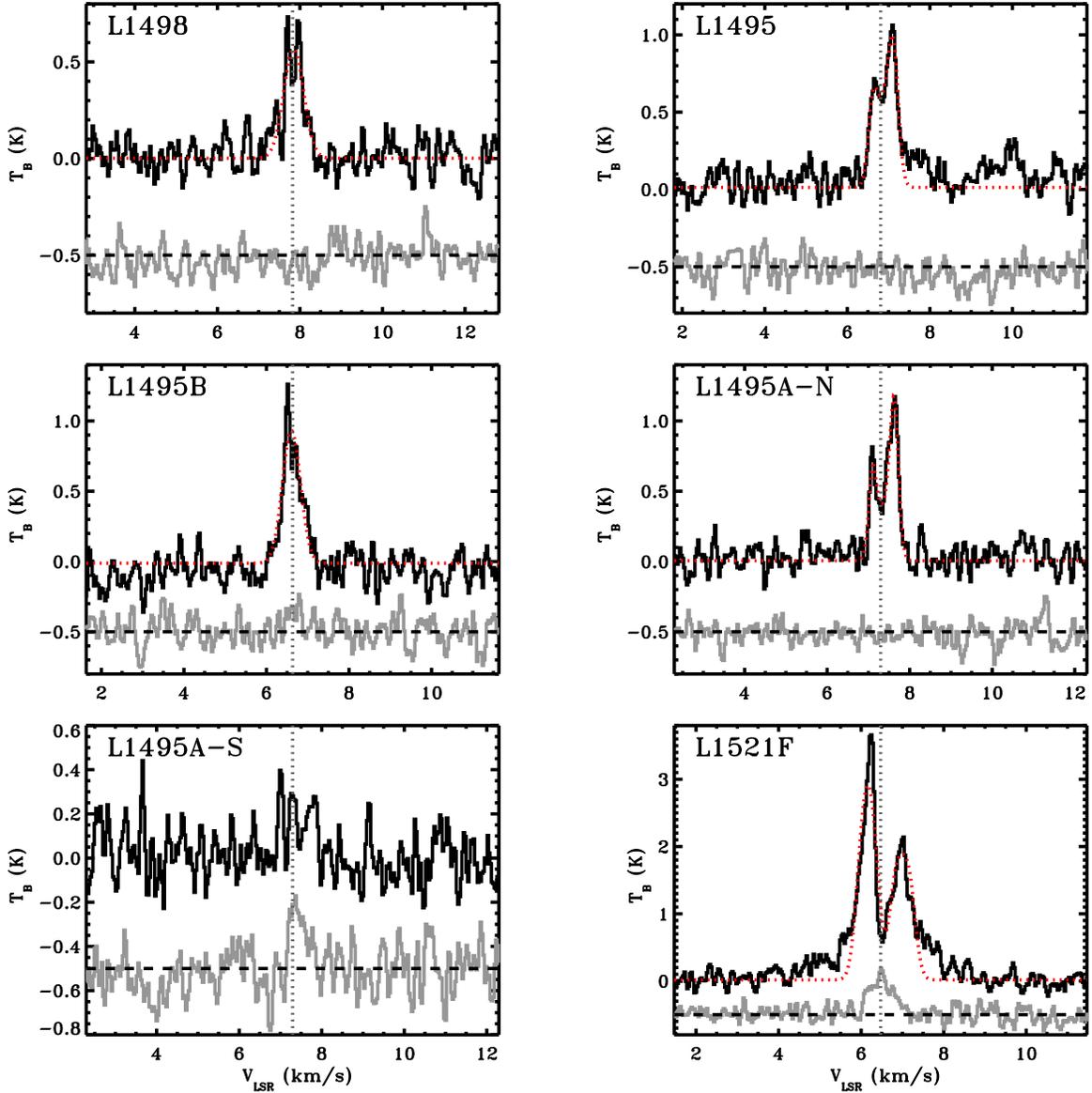}
\caption{Observed \HCOP\ (3-2) spectra (black) and \HTCOP\ (3-2)
  spectra (grey) for the cores L1498, L1495, L1495B, L1495A-N,
  L1495A-S, and L1521F.  The dotted vertical line shows the LSR
  velocities of \NTHP\ (1-0) \citep{Crapsi05}.  The \HTCOP\ spectra
  are offset by -0.5\,K, with a dashed horizontal line showing the
  zero level of the \HTCOP\ spectra. The red dotted lines shows the
  best fit to the observed \HCOP\ (3-2) spectra.  \label{SPECTRA1A}}
\end{figure}

\begin{figure}
\epsscale{1.0} 
\plotone{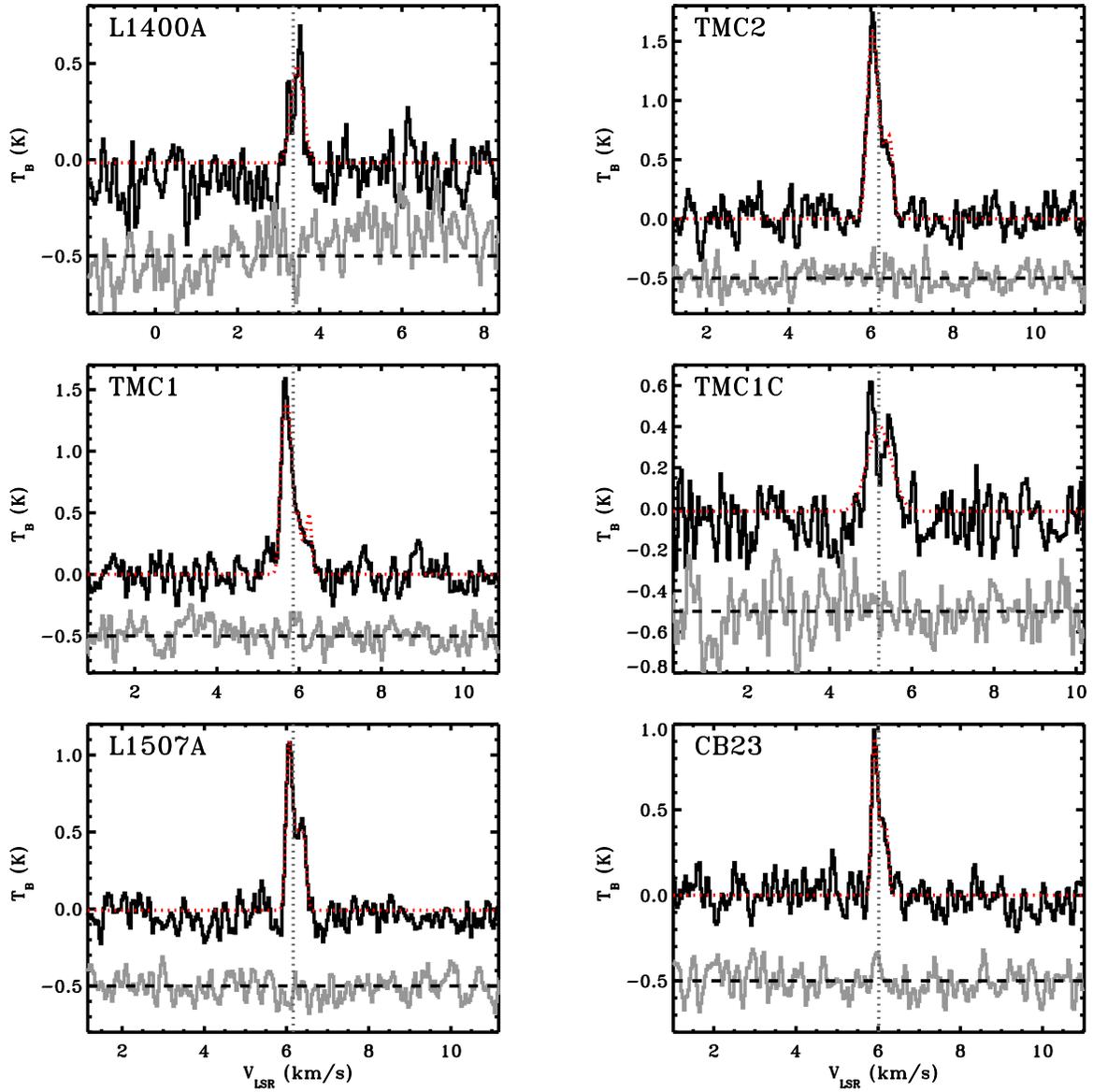}
\caption{Same as Figure \ref{SPECTRA1A}, for the cores L1400A, TMC2,
  TMC1, TMC1C, L1507A, and CB23. \label{SPECTRA1B}}
\end{figure}

\begin{figure}
\epsscale{1.0} 
\plotone{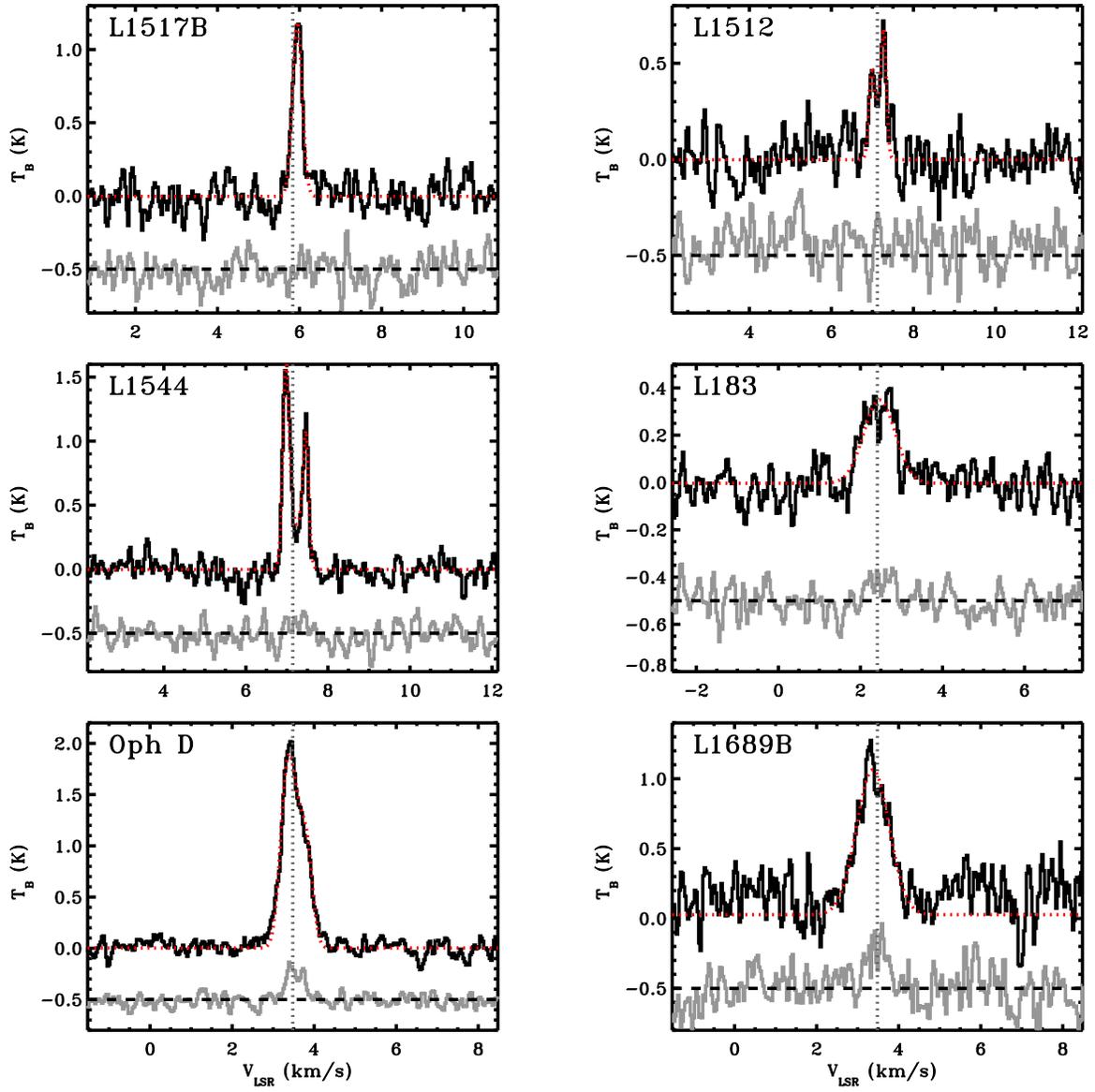}
\caption{Same as Figure \ref{SPECTRA1A}, for the cores L1517B, L1512,
  L1544, L183, Oph D, and L1689B. \label{SPECTRA1C}}
\end{figure}

\begin{figure}
\epsscale{1.0} 
\plotone{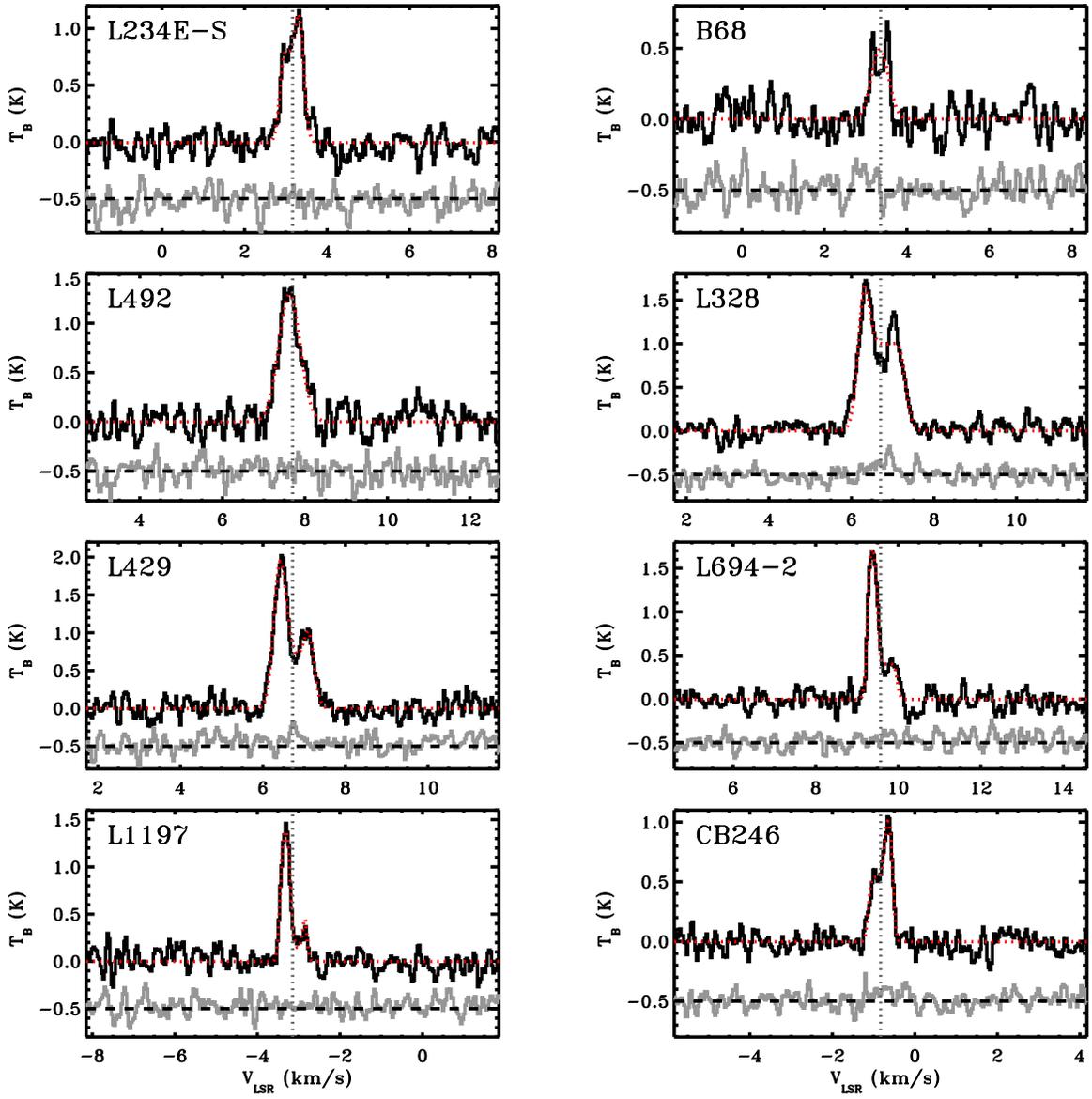}
\caption{Same as Figure \ref{SPECTRA1A}, for the cores L234E-S, B68,
  L492, L328, L429, L694-2, L1197, and CB 246. \label{SPECTRA1D}}
\end{figure}

\renewcommand{\thefigure}{\arabic{figure}}
\setcounter{figure}{1}
\begin{figure}
\epsscale{0.35} 
\plotone{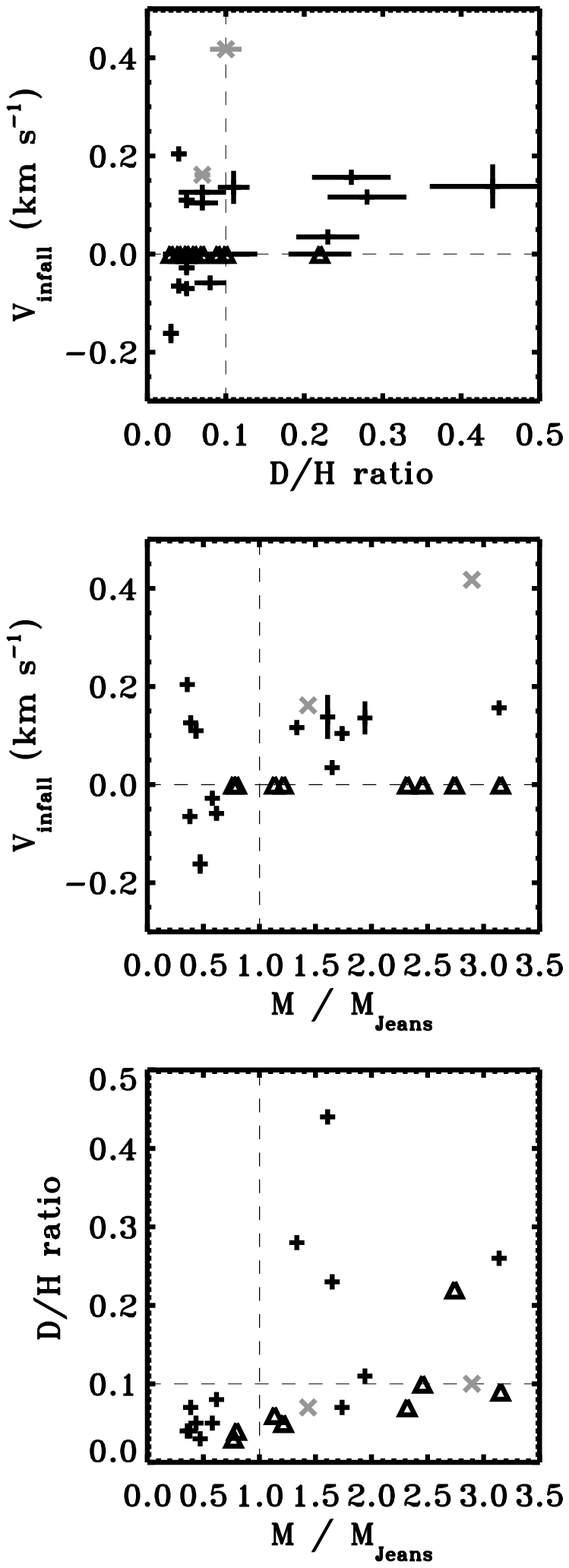}
\caption{ Infall velocity plotted against the $N$(\NTDP)/$N$(\NTHP)
  from \citet{Crapsi05} ({\it top}) and the ratio of the core mass to
  the thermal Jeans mass, as calculated from Equations \ref{MASSEQ}
  and \ref{JEANSEQ} and reported in Table \ref{PROPTABLE} ({\it
    middle}).  Positive values of $V_{\rm infall}$ indicate inward
  motions.  The $N$(\NTDP)/$N$(\NTHP) ratio against the
  $M/$\JEANS\ ratio is plotted in the bottom panel.  The protostellar
  cores L328 and L1521F are represented by grey 'x' symbols and all of
  the other cores are thought to be starless.  ``Static cores'' are
  given an infall/outflow velocity of 0\,\kms and are represented by
  triangles.  The cores L1495 and L1400A are not plotted in the middle
  and bottom panels because we could not determine $M/$\JEANS\ for
  them from the literature.  The core L1495A-S is not plotted in any
  panel because we did not detect it in our \HCOP\ observations.
\label{INFALLFIG}}
\end{figure}

\begin{deluxetable}{ccccccc} 
\tablewidth{0pt}
\tabletypesize{\scriptsize}
\tablecaption{Summary of \HCOP\ Observations \label{OBSTABLE}}
\tablehead{
 \colhead{Name}                 & 
 \colhead{RA\tablenotemark{1}}  &
 \colhead{Dec\tablenotemark{1}} &
 \colhead{Peak Intensity}       &
 \colhead{Integrated Intensity} &
 \colhead{Velocity Dispersion}  &
 \colhead{rms}                  \\
 \colhead{}                     &
 \colhead{J2000}                &
 \colhead{J2000}                &
 \colhead{T$_{\rm mb}$ (K)}        &
 \colhead{T$_{\rm mb}$ (K\,\kms)}  &
 \colhead{\kms}                 &
 \colhead{T$_{\rm mb}$ (K)}}                  
\startdata
  L1498                   & 04:10:51.5 & +25:09:58 & 0.57 & 0.34 & 0.24 & 0.08\\
  L1495                   & 04:14:11.2 & +28:08:56 & 1.00 & 0.58 & 0.16 & 0.09\\
  L1495B\tablenotemark{2} & 04:18:05.1 & +28:22:22 & 0.93 & 0.52 & 0.22 & 0.11\\
  L1495A-N                & 04:18:31.8 & +28:27:30 & 1.17 & 0.57 & 0.14 & 0.09\\
  L1495A-S                & 04:18:41.8 & +28:23:50 & N/A  & N/A  & N/A  & 0.10\\ 
  L1521F\tablenotemark{3} & 04:28:39.1 & +26:51:35 & 2.91 & 2.73 & 0.19 & 0.10\\
  L1400A                  & 04:30:56.8 & +54:52:36 & 0.49 & 0.19 & 0.15 & 0.14\\
  TMC 2                   & 04:32:48.7 & +24:25:52 & 1.59 & 0.71 & 0.11 & 0.11\\
  TMC 1                   & 04:41:32.9 & +25:44:44 & 1.39 & 0.61 & 0.09 & 0.11\\
  TMC 1C                  & 04:41:38.8 & +26:00:22 & 0.40 & 0.30 & 0.29 & 0.11\\
  L1507A                  & 04:42:38.6 & +29:43:45 & 1.10 & 0.37 & 0.08 & 0.08\\
  CB 23                   & 04:43:27.7 & +29:39:11 & 0.92 & 0.26 & 0.07 & 0.07\\
  L1517B                  & 04:55:17.6 & +30:37:49 & 1.19 & 0.35 & 0.12 & 0.09\\
  L1512                   & 05:04:09.7 & +32:43:09 & 0.68 & 0.22 & 0.09 & 0.10\\
  L1544                   & 05:04:16.6 & +25:10:48 & 1.66 & 0.61 & 0.12 & 0.08\\
  L183                    & 15:54:08.4 & -02:52:23 & 0.35 & 0.34 & 0.38 & 0.06\\
  Oph D                   & 16:28:28.9 & -24:19:19 & 1.90 & 1.29 & 0.17 & 0.06\\
  L1689B                  & 16:34:45.8 & -24:37:50 & 1.06 & 1.04 & 0.40 & 0.16\\
  L234E-S                 & 16:48:08.6 & -10:57:25 & 1.13 & 0.59 & 0.14 & 0.09\\
  B68                     & 17:22:38.9 & -23:49:46 & 0.49 & 0.26 & 0.21 & 0.09\\
  L492                    & 18:15:47.4 & -03:45:53 & 1.29 & 0.84 & 0.26 & 0.13\\
  L328\tablenotemark{4}   & 18:17:00.4 & -18:01:52 & 1.65 & 1.43 & 0.22 & 0.09\\
  L429                    & 18:17:05.1 & -08:13:40 & 1.94 & 1.30 & 0.19 & 0.10\\
  L694-2                  & 19:41:04.5 & +10:57:02 & 1.72 & 0.63 & 0.10 & 0.10\\
  L1197                   & 22:37:02.3 & +58:57:21 & 1.38 & 0.46 & 0.09 & 0.10\\
  CB 246                  & 23:56:41.5 & +58:34:09 & 1.00 & 0.44 & 0.10 & 0.07

\enddata
\tablenotetext{1}{From \citet{Crapsi05}}
\tablenotetext{2}{Name and coordinates from \citet{LM99}, not to be
  confused with LDN 1495B with J2000 coordinates 04:15:36.5 +28:46:06}
\tablenotetext{3}{Protostellar \citep{Bourke06}}
\tablenotetext{4}{Protostellar \citep{Lee09}}
\end{deluxetable}

\begin{deluxetable}{cccccccc} 
\tablewidth{0pt}
\tabletypesize{\scriptsize}
\tablecaption{Summary of Properties from Literature\label{PROPTABLE}}
\tablehead{
 \colhead{Name}                  &  
 \colhead{Distance}              &
 \colhead{N(\NTDP)/N(\NTHP)\tablenotemark{1}}     &
 \colhead{Radius}                &
 \colhead{Radius}                &
 \colhead{Mass}                  &
 \colhead{\JEANS}                &
 \colhead{M/\JEANS}              \\
 \colhead{}                      &
 \colhead{pc}                    &
 \colhead{}                      &
 \colhead{\arcsec}               &
 \colhead{pc}                    &
 \colhead{\MSUN}                 &
 \colhead{\MSUN}                 &
 \colhead{}}
\startdata
  L1498    & 140\tablenotemark{1} & 0.04 $\pm$ 0.01 &  56.0\tablenotemark{3} & 0.038 & 0.8\tablenotemark{3} & 1.0 & 0.8 \\
  L1495    & 140\tablenotemark{1} & 0.05 $\pm$ 0.01 & ...  \tablenotemark{4} & ...   & ...\tablenotemark{4} & ... & ... \\ 
  L1495B   & 140\tablenotemark{1} & 0.10 $\pm$ 0.04 &  92.8\tablenotemark{5} & 0.063 & 4.2\tablenotemark{5} & 1.7 & 2.5 \\
  L1495A-N & 140\tablenotemark{1} & 0.04 $\pm$ 0.01 &  15.5\tablenotemark{3} & 0.010 & 0.1\tablenotemark{3} & 0.3 & 0.3 \\ 
  L1495A-S & 140\tablenotemark{1} & 0.08 $\pm$ 0.03 &  20.9\tablenotemark{3} & 0.014 & 0.7\tablenotemark{3} & 0.4 & 1.7 \\
  L1521F\tablenotemark{7} & 140\tablenotemark{1} & 0.10 $\pm$ 0.02 &  83.0\tablenotemark{3} & 0.056 & 4.4\tablenotemark{3} & 1.5 & 2.9 \\ 
  L1400A   & 140\tablenotemark{1} & 0.05 $\pm$ 0.02 & ...  \tablenotemark{4} & ...   & ...\tablenotemark{4} & ... & ... \\
  TMC 2    & 140\tablenotemark{1} & 0.11 $\pm$ 0.02 & 230.6\tablenotemark{2} & 0.156 & 8.2\tablenotemark{2} & 4.2 & 2.0 \\
  TMC 1    & 140\tablenotemark{1} & 0.04 $\pm$ 0.01 &  31.9\tablenotemark{3} & 0.022 & 0.2\tablenotemark{3} & 0.6 & 0.3 \\
  TMC 1C   & 140\tablenotemark{1} & 0.07 $\pm$ 0.02 &  87.1\tablenotemark{3} & 0.059 & 3.7\tablenotemark{3} & 1.6 & 2.3 \\
  L1507A   & 140\tablenotemark{1} & 0.05 $\pm$ 0.01 &  45.2\tablenotemark{3} & 0.031 & 0.4\tablenotemark{3} & 0.8 & 0.5 \\
  CB 23    & 140\tablenotemark{1} & 0.07 $\pm$ 0.02 &  78.0\tablenotemark{6} & 0.053 & 2.5\tablenotemark{6} & 1.4 & 1.8 \\
  L1517B   & 140\tablenotemark{1} & 0.06 $\pm$ 0.01 &  77.1\tablenotemark{3} & 0.052 & 1.6\tablenotemark{3} & 1.4 & 1.1 \\
  L1512    & 140\tablenotemark{1} & 0.05 $\pm$ 0.01 &  47.5\tablenotemark{3} & 0.032 & 0.5\tablenotemark{3} & 0.9 & 0.6 \\
  L1544    & 140\tablenotemark{1} & 0.23 $\pm$ 0.04 &  60.1\tablenotemark{3} & 0.041 & 1.8\tablenotemark{3} & 1.1 & 1.6 \\
  L183     & 165\tablenotemark{1} & 0.22 $\pm$ 0.04 &  71.3\tablenotemark{3} & 0.057 & 4.3\tablenotemark{3} & 1.6 & 2.7 \\
  Oph D    & 165\tablenotemark{1} & 0.44 $\pm$ 0.08 &  55.3\tablenotemark{3} & 0.055 & 1.9\tablenotemark{3} & 1.2 & 1.6 \\
  L1689B   & 165\tablenotemark{1} & 0.09 $\pm$ 0.04 &  78.4\tablenotemark{3} & 0.063 & 5.4\tablenotemark{3} & 1.7 & 3.2 \\
  L234E-S  & 165\tablenotemark{1} & 0.08 $\pm$ 0.02 &  27.7\tablenotemark{3} & 0.022 & 0.4\tablenotemark{3} & 0.6 & 0.7 \\
  B68      & 125\tablenotemark{1} & 0.03 $\pm$ 0.01 &  57.5\tablenotemark{3} & 0.035 & 0.7\tablenotemark{3} & 1.0 & 0.7 \\
  L492     & 200\tablenotemark{1} & 0.05 $\pm$ 0.01 & 161.1\tablenotemark{2} & 0.156 & 5.2\tablenotemark{2} & 4.2 & 1.2 \\
  L328\tablenotemark{8} & 200\tablenotemark{1} & 0.07 $\pm$ 0.01 &  31.7\tablenotemark{3} & 0.031 & 1.2\tablenotemark{3} & 0.8 & 1.5 \\
  L429     & 200\tablenotemark{1} & 0.28 $\pm$ 0.05 &  19.4\tablenotemark{3} & 0.019 & 0.7\tablenotemark{3} & 0.5 & 1.4 \\
  L694-2   & 250\tablenotemark{1} & 0.26 $\pm$ 0.05 &  66.4\tablenotemark{3} & 0.080 & 6.8\tablenotemark{3} & 2.2 & 3.1 \\
  L1197    & 140\tablenotemark{3} & 0.07 $\pm$ 0.03 &  19.1\tablenotemark{3} & 0.013 & 0.3\tablenotemark{3} & 0.8 & 0.4 \\
  CB 246   & 140\tablenotemark{1} & 0.03 $\pm$ 0.01 &  27.9\tablenotemark{3} & 0.019 & 0.2\tablenotemark{3} & 0.5 & 0.4
\enddata
\tablenotetext{1}{\citep{Crapsi05}}
\tablenotetext{2}{\citep{Kauffmann08}}
\tablenotetext{3}{\citep{DiFrancesco08}}
\tablenotetext{4}{Core mass and radius not found in the literature}
\tablenotetext{5}{\citep{Hirota04}}
\tablenotetext{6}{\citep{Stutz09}}
\tablenotetext{7}{Protostellar \citep{Bourke06}}
\tablenotetext{8}{Protostellar \citep{Lee09}}\end{deluxetable}

\setlength{\tabcolsep}{0.01in}
\begin{deluxetable}{llllllll}
\tablewidth{0pt}
\tabletypesize{\tiny}
\tablecaption{Summary of Velocity Information\label{VELOCITYTABLE}}
\tablehead{
 \colhead{Name}                  & 
 \colhead{\HCOP\ (3-2)}          &
 \colhead{Model Type}            &
 \colhead{HCN (1-0) }            &
 \colhead{CS, HCN, and \NTHP\ }  &
 \colhead{\NTHP\ (1-0) }         &
 \colhead{\HCOP\ (3-2)}          &
 \colhead{\NTHP\ (1-0)}          \\
 \colhead{}                                  &
 \colhead{\VINOUT \tablenotemark{5}}  &
 \colhead{}                                  &
 \colhead{Classification}                    &
 \colhead{Classification}                    &
 \colhead{Skewness \tablenotemark{5}} &
 \colhead{V$_{\rm LSR}$}                 &
 \colhead{V$_{\rm LSR}$}                 \\
 \colhead{} & 
 \colhead{(\kms)} &
 \colhead{} &
 \colhead{} & 
 \colhead{} &  
 \colhead{(\kms)} &
 \colhead{(\kms)} &
 \colhead{(\kms)} \\
 \colhead{}                     &
 \colhead{This work}            &
 \colhead{}                     &
 \colhead{\citep{Sohn07}}       &
 \colhead{\citep{Lee11}}        &
 \colhead{\citep{Crapsi05}}     &
 \colhead{This work}            &
 \colhead{\citep{Crapsi05}}}
\startdata
  L1498                   & static core      & Gaussian  & ``contracting core''\tablenotemark{2} & ``contracting core'' & -0.01 $\pm$ 0.05 &  7.85 $\pm$ 0.01 &  7.822 $\pm$ 0.001 \\
  L1495                   & -0.07 $\pm$ 0.01 & Hill 5    & ...                                   & ...                  &  0.10 $\pm$ 0.05 &  6.89 $\pm$ 0.01 &  6.807 $\pm$ 0.001 \\ 
  L1495B                  & static core      & Gaussian  & ...                                   & ``static core''      & ...              &  6.68 $\pm$ 0.01 &  6.633 $\pm$ 0.008 \\ 
  L1495A-N                & -0.07 $\pm$ 0.01 & Hill 5    & ``contracting core''\tablenotemark{1} & ``oscillating core'' & -0.04 $\pm$ 0.05 &  7.40 $\pm$ 0.01 &  7.296 $\pm$ 0.003 \\ 
  L1495A-S                & non-detection    & ...       & ``contracting core''\tablenotemark{2} & ``contracting core'' & ...              &  non-detection   &  7.294 $\pm$ 0.007 \\ 
  L1521F\tablenotemark{6} &  0.42 $\pm$ 0.01 & Hill 5    & ``contracting core''\tablenotemark{1} & ``expanding core''   & 0.19 $\pm$ 0.06  &  6.58 $\pm$ 0.01 &  6.472 $\pm$ 0.001 \\ 
  L1400A                  & static core      & Gaussian  & ``expanding core''\tablenotemark{3}   & ``static core''      & ...              &  3.40 $\pm$ 0.01 &  3.355 $\pm$ 0.002 \\ 
  TMC 2                   &  0.14 $\pm$ 0.03 & Hill 5    & ``contracting core''\tablenotemark{2} & ``contracting core'' & -0.01 $\pm$ 0.05 &  6.20 $\pm$ 0.01 &  6.193 $\pm$ 0.001 \\ 
  TMC 1                   &  0.20 $\pm$ 0.01 & Hill 5    & ``contracting core''\tablenotemark{2} & ``contracting core'' & -0.13 $\pm$ 0.06 &  5.90 $\pm$ 0.01 &  5.856 $\pm$ 0.003 \\ 
  TMC 1C                  & static core      & Gaussian  & ...                                   & ...                  &  0.43 $\pm$ 0.10 &  5.24 $\pm$ 0.01 &  5.196 $\pm$ 0.003 \\ 
  L1507A                  &  0.11 $\pm$ 0.02 & Two Layer & ``contracting core''\tablenotemark{2} &``oscillating core''  & ...              &  6.35 $\pm$ 0.01 &  6.163 $\pm$ 0.004 \\ 
  CB 23                   &  0.10 $\pm$ 0.01 & Two Layer & ``contracting core''\tablenotemark{2} & ``static core''      & ...              &  6.15 $\pm$ 0.01 &  6.015 $\pm$ 0.002 \\ 
  L1517B                  &  static core     & Gaussian  & ``expanding core''\tablenotemark{3}   & ``static core''      & -0.08 $\pm$ 0.05 &  5.95 $\pm$ 0.01 &  5.835 $\pm$ 0.001 \\ 
  L1512                   & -0.03 $\pm$ 0.01 & Hill 5    & ``expanding core''\tablenotemark{3}   & ``oscillating core'' &  0.02 $\pm$ 0.05 &  7.13 $\pm$ 0.01 &  7.121 $\pm$ 0.001 \\ 
  L1544                   &  0.03 $\pm$ 0.01 & Hill 5    & ``contracting core''\tablenotemark{1} & ``contracting core'' &  0.40 $\pm$ 0.09 &  7.21 $\pm$ 0.01 &  7.143 $\pm$ 0.002 \\ 
  L183                    & static core      & Gaussian  & ``oscillating core''\tablenotemark{4} & ``contracting core'' &  0.09 $\pm$ 0.05 &  2.22 $\pm$ 0.09 &  2.413 $\pm$ 0.001 \\ 
  Oph D                   &  0.14 $\pm$ 0.04 & Hill 5    & ``contracting core''\tablenotemark{2} & ``static core''      &  0.54 $\pm$ 0.12 &  3.54 $\pm$ 0.01 &  3.478 $\pm$ 0.001 \\ 
  L1689B                  & static core      & Gaussian  & ``contracting core''\tablenotemark{1} & ``contracting core'' & ...              &  3.37 $\pm$ 0.01 &  3.481 $\pm$ 0.005 \\ 
  L234E-S                 & -0.06 $\pm$ 0.01 & Hill 5    & ``oscillating core''\tablenotemark{4} & ``contracting core'' & -0.25 $\pm$ 0.07 &  3.16 $\pm$ 0.01 &  3.164 $\pm$ 0.003 \\ 
  B68                     & static core      & Gaussian  & ...                                   & ...                  & -0.09 $\pm$ 0.05 &  3.35 $\pm$ 0.02 &  3.364 $\pm$ 0.001 \\ 
  L492                    & static core      & Gaussian  & ``contracting core''\tablenotemark{1} & ``contracting core'' &  0.25 $\pm$ 0.07 &  7.63 $\pm$ 0.01 &  7.701 $\pm$ 0.001 \\ 
  L328\tablenotemark{7}   &  0.16 $\pm$ 0.01 & Two Layer & ...                                   & ...                  & -0.02 $\pm$ 0.08 &  6.91 $\pm$ 0.01 &  6.707 $\pm$ 0.002 \\ 
  L429                    &  0.12 $\pm$ 0.01 & Hill 5    & ``oscillating core''\tablenotemark{4} & ``expanding core''   & -0.20 $\pm$ 0.10 &  6.72 $\pm$ 0.01 &  6.719 $\pm$ 0.001 \\ 
  L694-2                  &  0.16 $\pm$ 0.01 & Two Layer & ``contracting core''\tablenotemark{1} & ``contracting core'' &  0.22 $\pm$ 0.07 &  9.77 $\pm$ 0.01 &  9.574 $\pm$ 0.001 \\ 
  L1197                   &  0.13 $\pm$ 0.01 & Hill 5    & ``contracting core''\tablenotemark{2} & ``contracting core'' &  0.08 $\pm$ 0.11 & -3.14 $\pm$ 0.01 & -3.147 $\pm$ 0.002 \\ 
  CB 246                  & -0.16 $\pm$ 0.02 & Two Layer & ``oscillating core''\tablenotemark{4} & ``expanding core''   &  0.02 $\pm$ 0.05 & -0.99 $\pm$ 0.02 & -0.830 $\pm$ 0.001 \\
\enddata  
\tablenotetext{1}{All three hyperfine components of HCN (1-0) show infall}
\tablenotetext{2}{Some hyperfine components of HCN (1-0) show infall}
\tablenotetext{3}{Some hyperfine components of HCN (1-0) show outflow}
\tablenotetext{4}{At least one of the three hyperfine components of
  HCN (1-0) shows inward motions and at least one of the three
  hyperfine components shows outward motions}
\tablenotetext{5}{Positive values indicate inward motion, negative values indicate outward motion}
\tablenotetext{6}{Protostellar \citep{Bourke06}}
\tablenotetext{7}{Protostellar \citep{Lee09}}
\end{deluxetable}

\end{document}